# Role of Self-Assembled Monolayers on Improved Electrical Stability of Amorphous In-Ga-Zn-O Thin-Film Transistors

Xiaosong Du,[1] Brendan T. Flynn,[1] Joshua R. Motley,[1] William F. Stickle,[2] Hendrik Bluhm,[3] Gregory S. Herman[1,a]

[1] Oregon State University, School of Chemical, Biological, and Environmental Engineering, Corvallis, Oregon, 97331, USA

[2] Hewlett Packard Company, 1000 Northeast Circle Boulevard, Corvallis, Oregon 97330, USA

[3] Lawrence Berkeley National Laboratory, Chemical Sciences Division, One Cyclotron Road, Berkeley, CA 94720, USA

[a] Author to whom correspondence should be addressed. Electronic mail: greg.herman@oregonstate.edu

KEYWORDS: thin film transistor, back channel passivation, bias stress test.

# ABSTRACT

Self-assembled monolayers (SAMs) have been used to improve both the positive and negative bias-stress stability of amorphous indium gallium zinc oxide (IGZO) bottom gate thin film transistors (TFTs). N-hexylphosphonic acid (HPA) and fluorinated hexylphosphonic acid (FPA) SAMs adsorbed on IGZO back channel surfaces were shown to significantly reduce bias-stress turn-on voltage shifts compared to IGZO back channel surfaces with no SAMs. FPA was found to have a lower surface energy and lower packing density than HPA, as well as lower bias-stress turn-on voltage shifts. The improved stability of IGZO TFTs with SAMs can be primarily attributed to a reduction in molecular adsorption of contaminants on the IGZO back channel surface and minimal trapping states present with phosphonic acid binding to the IGZO surface.

Indium gallium zinc oxide (IGZO) is a very promising active material for thin film transistors (TFTs), especially when incorporated into flat panel and flexible displays.[1-4] IGZO-TFTs have relatively high average electron mobilities ($\mu_{avg}$>10 cm$^2$/Vs) despite being amorphous, and have been demonstrated to have excellent performance even after low temperature processing that is compatible with polymeric substrates.[1,5] IGZO-TFTs may eventually replace amorphous silicon TFT technologies for many display applications.[2,3] High electrical stability is an important requirement for display applications, since bias, current, and/or light induced instabilities may cause non-uniformities in relative brightness of individual pixels. High electrical stability is also critical for IGZO field effect sensors used for detecting gases,[6] temperature,[7] light,[8] or for deoxyribonucleic acid.[9] For these reasons the bias stress stability of IGZO-TFTs has been intensely investigated where effects of composition, process temperature, process conditions, gate dielectric, back channel passivation, and many other factors have been evaluated.[10-15]

Proposed mechanisms for bias stress threshold voltage ($V_{th}$) shifts include electron trapping at the semiconductor/dielectric interface, electron injection into the dielectric, or formation of sub-band gap states in the bulk of films. It has also been suggested that adsorbed species (e.g., $O_2$ and $H_2O$) on the back channel of the IGZO-TFTs can be the dominant mechanism for device instability, by providing acceptor or donor states or through field induced adsorption or desorption of these species.[17-20] For example, molecular oxygen can form a depletion layer at the back channel surface through electron acceptor states, while absorbed water can form an accumulation layer at the back channel surface through electron donor states. A depletion layer results in a positive $V_{th}$ shift, whereas an accumulation layer results in a negative $V_{th}$ shift. Recently it has been demonstrated that dense passivation layers[21] or self-assembled monolayers (SAMs)[22] on IGZO-TFT back channels can significantly minimize these threshold voltage shifts. A potential benefit of using SAMs for surface passivation is that one can obtain well controlled chemistries at the back channel interface, even when using polymeric dielectrics for flexible electronic applications or integrating polymeric sensing layers.

Herein we investigate the use of n-hexylphosphonic acid (n-HPA) and (3,3,4,4,5,5,6,6,6-nonaflurohexyl)phosphonic acid (FPA) self-assembled monolayers (SAMs) as molecular passivation layers for IGZO-TFT back channel surfaces. The goal of these studies is to evaluate the effects of surface energy and SAM molecular dipole on device performance. IGZO surface chemistries were characterized using contact angle measurements, positive and negative bias stress turn-on voltage shifts, and X-ray photoelectron spectroscopy (XPS). Significant improvements in IGZO-TFT turn-on voltage ($V_{ON}$) stabilities were observed for both the HPA and FPA passivated surfaces compared to the unpassivated surfaces. These results provide insight into the role of defect passivation at back channel surfaces of IGZO and other transparent amorphous oxide semiconductors (TAOS).

IGZO-TFT test structures were fabricated using a heavily p-doped Si substrate as the gate and thermally grown $SiO_2$ (100 nm) as the gate dielectric. Substrates were cleaned prior to deposition of IGZO with acetone, isopropyl alcohol, and DI water. Amorphous IGZO films (~50 nm thick) were deposited using RF magnetron sputter deposition with a 3 in. IGZO target (molar composition: $In_2O_3:Ga_2O_3:ZnO$), 100W RF power, ~4 mTorr chamber pressure, and 20 sccm flow rate with a 1:19 ($O_2$:Ar) ratio. IGZO active layers were patterned using a shadow mask during deposition, and the films were subsequently annealed in air to 300 °C. Source and drain electrodes were patterned using a shadow mask during thermal evaporation of Al (~500 nm thick) giving a W/L ratio of 1000 μm/100 μm. Back channel surface passivation using SAMs was performed by UV-Ozone treatment for 5 minutes followed by immediate immersion of the IGZO-TFTs in 2mM n-HPA or FPA solution in 95% ethanol. Samples were removed from the solution, rinsed with 95% ethanol, and then dried with flowing nitrogen. The water contact angle was found to increase for increasing soak times due to adsorption of n-HPA or FPA molecules and a corresponding reduction in surface energy as a result of the alkane or fluorinated alkane groups on the SAMs. The highest contact angles for n-HPA and FPA treated IGZO surfaces, after a 24 h soak, was 84° and 94°, respectively, compared to 23° which was obtained for an IGZO surface after only the UV/Ozone treatment. A 24 h soak was used for all the IGZO-TFTs in this study. In order to remove physisorbed SAMs the samples were further sonicated in a 5% triethylamine/ethanol solution, rinsed with ethanol, and then dried with nitrogen. Duplicate IGZO-TFTs were fabricated where they have gone through identical processing steps as the passivated samples, other than n-HPA or FPA soak. A schematic of the device structure and possible bonding of n-HPA or FPA to the IGZO surface is shown in Figure 1.

All IGZO-TFT bias-stress measurements were performed in the dark at room temperature in air using an Agilent 4155C precision semiconductor parameter analyzer. Single sweep drain-to-source current versus gate voltage ($I_{DS}$-$V_{GS}$) transfer curves were measured with the drain voltage ($V_{DS}$) set to 1 V. Stress measurements were performed up to $10^4$ s using an applied $V_{GS}$ of ±10 V (1 MV/cm) with $V_{DS}$ set to 0 V. Synchrotron-based XPS measurements were performed using a specially designed endstation at the undulator beamline 11.0.2 at the Advanced Light Source.[23] A photon energy of hν = 735 eV was used to obtain surface sensitive spectra. The data were acquired with a 90º emission angle and an electron analyzer pass energy of 10 eV. Laboratory-based XPS measurements were performed with a PHI Quantera Scanning ESCA system using monochromatized Al Kα radiation with a 200 μm spot size. The data were acquired with a 45º emission angle and an electron analyzer pass energy of 69 eV. The XPS data were fit using CasaXPS, where the most intense peak in the spectrum was used to define the core-level full-width-half-maximum (FWHM) and Gaussian-Lorentzian mixing. A linear background was used to fit all spectra. The binding energies were charge corrected to the C 1s aliphatic carbon peak at 284.6 eV.

In Table 1 we provide $\mu_{avg}$, $V_{ON}$, drain current on-to-off ratio ($I_{ON}/I_{OFF}$), and hysteresis (difference in the up and down sweeps in the $I_{DS}$-$V_{GS}$ data) for unstressed devices with different surface treatments. Both $\mu_{avg}$ and $V_{ON}$ were extracted using methods described previously.[24] Only very small changes were observed in $\mu_{avg}$, $V_{ON}$, and $I_{ON}/I_{OFF}$, irrespective of surface treatment, where the major difference is a decrease in $V_{ON}$ and a reduction in hysteresis with adsorption of SAMs on the back channel surface. In Figure 2, we show positive and negative bias stress $I_{DS}$-$V_{GS}$ data for IGZO-TFTs without and with n-HPA or FPA passivation for bias stress times of 0, 10, 100, 1,000, and 10,000 s. The arrows in the figure indicate the direction of turn on voltage shift ($\Delta V_{ON}$) for increasing bias stress time. It was observed that as the duration of bias stress increased the $\Delta V_{ON}$ continued to shift. However, we found that both sub-threshold slope and $\mu_{avg}$ remained unchanged after either positive or negative bias stress.

In Figure 3, we plot $\Delta V_{ON}$ versus bias stress time for the IGZO-TFTs. The experimental data in the figure is given as open and closed symbols, which represent positive and negative bias stress measurements, respectively. Devices without passivation had the largest bias stress $\Delta V_{ON}$, where the devices had shifts of 7.2 and -8.0 V after 10,000 s of positive and negative bias stress, respectively. For bias stress induced trapping and meta-stability, a stretched exponential model has been previously employed for amorphous silicon-[25] and IGZO-TFTs.[15,18,20,21]

$$\Delta V_{th} = \Delta V_O \left\{ 1 - \exp\left[ -\left(\frac{t}{\tau}\right)^\beta \right] \right\} \qquad (1)$$

The threshold voltage shift ($\Delta V_{th}$) is related to the effective voltage drop across the gate oxide ($\Delta V_O$), which is approximately given by the difference between gate stress voltage and initial threshold voltage ($V_{G,stress}-V_{th,i}$), the characteristic trapping time of carriers ($\tau$), and stretched exponential exponent ($\beta$). Both $\tau$ and $\beta$ are dependent on stress bias, materials, and device properties. For more accuracy we used $V_{ON}$ in place of $V_{th}$, where $V_{ON}$ was obtained by extrapolation from log $I_{DS}$ versus $V_{GS}$ plots where log ($I_{DS}/(W/L)$) (A) is equal to -11. The stretched exponential model for IGZO devices provided excellent fits to bias stress $V_{ON}$ shifts for devices without passivation (dashed line through the circle data points in Figure 3). Values of $\tau$ and $\beta$ for different surface treatments are summarized in Table 2, and were consistent with literature values. For example, characteristic trapping times for IGZO TFTs in the literature are between ~$10^3$ to $10^6$ s,[15,18,20,21] and can vary substantially depending on whether measuring negative or positive bias stress, the gate dielectric used, and the use of passivation layers to prevent chemistry induced trapping at the back channel interface. The stretched exponential model is primarily used to describe $V_{ON}$ shifts due to trapping at the dielectric interface or due to metastability in the channel material, although this model has also been used to describe $V_{ON}$ shifts due to back channel chemical effects for IGZO[20] and zinc tin oxide (ZTO)[26] TFTs. In Figure 3 we show the bias stress results for an IGZO-TFT with n-HPA adsorbed on the surface (triangle data points). There is a significant reduction in bias stress shift, especially for negative biases, when compared to devices without SAMs on the back channel surface. Similar improvements in positive bias stress stability have previously been shown for ZTO-TFTs when n-HPA is adsorbed on the backchannel surface.[26] In Figure 3, we show that IGZO TFTs with FPA SAM had the lowest positive and negative bias stress shifts (square data points). The observed shifts for FPA passivated surfaces are significantly reduced compared to the other surface preparations, and we find that the stretched exponential model no longer provides an adequate fit to the data. However, we now find that a linear relationship between $\Delta V_{th}$ and logarithmic time provides a good fit to our data for FPA passivated IGZO TFTs. This relationship can be described as following the total charge that becomes trapped at the channel/dielectric interface or in the dielectric for any given time $t$.[12] The positive and negative stress test showed a slope of 0.35 and -0.23, respectively.

Since devices with and without SAM passivation had identical preparation of the dielectric/IGZO interface and the IGZO post-deposition annealing conditions we are able to make several observations. In the case of the unpassivated surface the bias stress $V_{ON}$ shifts were most likely due to chemistry at the back channel interface as opposed to trapping at the dielectric interface or due to metastability in the channel material. Since the trends for the n-HPA passivated IGZO TFTs are very similar to the unpassivated IGZO TFTs (both fit the stretched exponential model), it is likely that chemistry at the back channel interface still allow trap or donor states that lead to $V_{ON}$ shifts during bias stress. Finally, it appears that the FPA passivated IGZO TFT stabilities switch to a different mechanism (no longer fits the stretched exponential model), and in this case we believe that the device stability is no longer dictated by the back channel interface, but is now dominated by charge injection into the dielectric.

In the lower portion of Figure 4 we show surface-sensitive O 1s XPS data for an as deposited blanket IGZO film using a photon energy of hν = 735 eV. The sample was stored in ambient air for several days prior to analysis and had substantial carbon contamination (spectra not shown). To fit the O 1s core level spectrum it was necessary to use three components, similar to what has been done previously for IGZO.[27-29] From our analysis we have determined that the binding energies for the three components were at 530.1, 531.2, and 532.2 eV. For several prior IGZO XPS studies the three O 1s peak components were assigned as follows.[27,28] The low-energy peak corresponds to lattice oxygen ions, which have neighboring cations. The mid-energy O 1s peak corresponds to oxygen vacancies or more accurately, lattice oxygen ions in oxygen deficient regions of IGZO. The high-energy O 1s components were due to adsorbed species.

To assess these assignments we have performed *in-situ* annealing experiments with the goal to desorb impurities from the surface and/or increase the number of oxygen vacancies within the film. In the upper portion of Figure 4 we show surface-sensitive O 1s XPS data for the IGZO sample while annealing to 300 ºC in ultrahigh vacuum. During the anneal there was a significant reduction in the intensities of the mid- and high-energy O 1s components (quantitatively summarized in Table 3). Since the annealing was performed in an oxygen deficient environment the number of oxygen vacancies should increase, resulting in an increase in relative intensity for the mid-energy O 1s component based on the assignments given above.

However, we have found a significant decrease in intensity for both the mid- and high-energy O 1s components suggesting that both components are likely due to adsorbed species that desorb during the anneal. Our interpretation of the O 1s components follows the convention that the mid-energy O 1s peak corresponds to surface hydroxyls, and not oxygen deficient regions in the material. This interpretation of the O 1s spectra has been suggested by several other groups for IGZO as well.[29-31] Thus, the mid- and high-energy peak intensity is correlated with the number of adsorbed species on the IGZO back surface and the corresponding number of adsorbate-induced traps (i.e., not the number of oxygen vacancies in the IGZO film and the corresponding number of traps), and it is these adsorbed species that influence the bias stress stability of the devices fabricated using TAOS materials. Based on the bias stress data this would suggest that the FPA passivated surface has significantly fewer absorbed species on the IGZO back channel surface, which is consistent with the significant reduction in surface energy compared to the unpassivated and n-HPA passivated surfaces.

In the lower portion of Figure 5 we show O 1s XPS data for a blanket IGZO film after being annealed to 300 °C for 60 min in air, while in the upper portion of Figure 5 we show O 1s XPS data of samples with n-HPA and FPA to investigate changes in IGZO-TFT back channel chemistry. As mentioned above, the low-energy peak corresponds to lattice oxygen ions in IGZO. The n-HPA and FPA SAMs adsorbed on IGZO the mid-energy O 1s components can be attributed to P-O-In/P-O-Ga/P-O-Zn and P=O, which was consistent with prior studies.[32-34] The third peak at a higher binding energy of 532.3 eV can be assigned to surface contamination, as will be discussed below. The mid-energy peak intensity/low-energy peak intensity ratio ($I_{me}/I_{le}$) gives a qualitative estimate of the relative amount of SAMs adsorbed on the surface. As can be seen in Figure 5, and also given in Table 4, the $I_{me}/I_{le}$ intensity ratios are 70% and 49% for n-HPA and FPA, respectively. This suggests that there is more n-HPA adsorbed to the IGZO surface, which would be expected based on the relative sizes of the alkane and fluorinated alkane chains and the associated packing densities.[35]

In the lower portion of Fig. 6, we show C 1s spectra for a blanket IGZO film after being annealed to 300 °C. The C 1s spectra had three fairly well-separated peaks with the lower energy peak having an energy of 284.6 eV, which is assigned to aliphatic carbon (C–C or C–H bonds), and the two higher energy peaks at 286.2 eV and 288.8 eV can be assigned to C–O and O–C=O groups on the surface, respectively.[36,37] As shown in the middle portion of Fig. 6, IGZO sample with n-HPA has much less C–O and O–C=O groups on the surface, where most of the intensity comes from aliphatic carbon in HPA molecules, and two higher energy peaks at 286.4 and 288.5 eV due to –CH$_2$- bonded to the phosphonic acid group and surface contamination, respectively. In the upper portion of Fig. 6, we can see that the C 1s spectrum for FPA has five components, where the -CF$_2$- and -CF$_3$ groups in FPA molecules give peaks at 290.8 eV and 293.3 eV, respectively. The -CH$_2$- groups bound to -CF$_2$- and -PO$_3$ give peaks at 286.1 eV and 288.4 eV, respectively.

These results strongly suggest that the back channel surface chemistry was the dominant reason for bias-stress $V_{ON}$ shifts for the unpassivated and n-HPA passivated IGZO-TFTs. There are several mechanisms that have been proposed to describe device instabilities due to the back channel surface chemistry, including field-induced adsorption/desorption[16] and dipole interactions.[5,19] A significant difference between n-HPA and FPA is the orientation of the molecular dipole at the surface.[34] For example, n-HPA has a polar head group attached to the surface and a non-polar alkyl chain pointing away from the surface. FPA has a polar head group attached to the surface, but also has highly electronegative fluorine atoms on the alkyl chains pointing away from the surface. This results in a molecular dipole that points towards and away from the surface, respectively.

Recent studies suggested that water adsorption was the dominant species for ZTO TFT instabilities leading to significant shifts in threshold voltage,[38] and increased hysteresis in the TFT transfer characteristics.[39] The saturation coverage of SAMs used in these studies significantly limits the interaction of impurity species, especially $H_2O$, with the IGZO surface (i.e. it does not provide electron trap and/or donor states that would result in device instabilities). The FPA surface has a significantly lower surface energy compared to n-HPA, and as a result may further reduce the interaction of $H_2O$ with the IGZO back channel surface. This may help explain why the IGZO-TFT with FPA has lower $\Delta V_{ON}$ bias stress shifts, and lower device hysteresis than the n-HPA passivated devices, or the unpassivated devices. Finally, we have not observed significant differences in the IGZO TFT hysteresis for the different surface treatments, which suggests that SAMs do not degrade device performance.

In summary, highly stable IGZO TFTs were fabricated through molecular passivation of the oxide back channel surface. Both the positive and negative bias stress dependence of the turn-on voltage shifts for unpassivated IGZO-TFTs and n-HPA passivated IGZO-TFTs was best described by a stretched exponential model, whereas FPA passivated IGZO TFTs had a linear relationship between turn-on voltage shifts and logarithmic time for the same bias stress conditions. Bias stress measurements and XPS analysis suggest that reductions in the turn-on voltage shifts were due to well defined molecular passivation at the backchannel surface, which limits electrical modification of the channel due to the adsorption of impurity species (i.e., $O_2$, $H_2O$, etc.).

**Acknowledgements**

This work was funded in part by the Oregon Nanoscience and Microtechnologies Institute (ONAMI). X. Du acknowledges funding support from the Juvenile Diabetes Research Foundation (3-PDF-2014-113-A-N). The Advanced Light Source is supported by the Director, Office of Science, Office of Basic Energy Sciences, of the U.S. Department of Energy under Contract No. DE-AC02-05CH11231.

**Table 1.** Average mobility, turn on voltage, drain current on-to-off ratio and hysteresis for unstressed IGZO TFTs with/without n-HPA or FPA.

| Surface treatment | $\mu_{avg}$ (cm$^2$/Vs) | $V_{on}$ (V) | $I_{on}/I_{off}$ | Hysteresis (V) |
|---|---|---|---|---|
| Annealed | 20.2 | -0.7 | $5.0 \times 10^6$ | 0.6 |
| n-HPA | 20.6 | -1.4 | $3.0 \times 10^6$ | 0.4 |
| FPA | 21.4 | -2.1 | $4.0 \times 10^6$ | 0.3 |

**Table 2.** Characteristic trapping time of carriers, $\tau$ and stretched exponential exponent, $\beta$ for IGZO TFTs.

| Stress test | $\tau$(s) | $\beta$ |
|---|---|---|
| Annealed, positive stress | $1.0\times10^4$ | 0.45 |
| Annealed, negative stress | $5.5\times10^3$ | 0.74 |
| HPA, positive stress | $2.1\times10^5$ | 0.33 |
| HPA, negative stress | $7.7\times10^6$ | 0.22 |

**Table 3.** Surface-sensitive XPS binding energies and peak areas obtained from blanket IGZO films for as deposited and while annealing to 300 ºC in ultrahigh vacuum.

|  | O 1s B.E. (eV) | Area (%) |
|---|---|---|
| As deposited | 530.1 | 53.4 |
|  | 531.3 | 28.2 |
|  | 532.2 | 18.4 |
| 300 ºC | 530.1 | 88.7 |
|  | 531.3 | 11.3 |

**Table 4.** XPS binding energies and peak areas obtained from blanket IGZO films that have been annealed and treated with n-HPA or FPA.

|  | O 1s B.E. (eV) | Area (%) | C 1s B.E. (eV) | Area (%) |
|---|---|---|---|---|
| FPA |  |  | 284.6 | 46.1 |
|  | 530.1 | 61.9 | 286.1 | 11.8 |
|  | 531.2 | 30.5 | 288.4 | 9.5 |
|  | 532.3 | 7.6 | 290.8 | 22.7 |
|  |  |  | 293.3 | 9.9 |
| HPA | 530.1 | 53.3 | 284.6 | 90.6 |
|  | 531.2 | 37.4 | 286.4 | 6.9 |
|  | 532.3 | 9.3 | 288.5 | 2.5 |
| Annealed | 530.1 | 65.8 | 284.6 | 72.4 |
|  | 531.2 | 23.8 | 286.2 | 10.1 |
|  | 532.3 | 10.4 | 288.8 | 17.5 |

FIG. 1. Schematic illustration of SAM functionalized IGZO TFT structure showing the mono-, bi-, and tri-dentate binding of SAM to IGZO.

FIG. 2. Plot of $I_D$ vs $V_{GS}$ for various treatments of the IGZO TFTs as a function of positive and negative biased stress time of 0, 10, 100, 1,000 and 10,000 s. (a) Annealed sample, positive bias, (b) annealed sample, negative bias, (c) Annealed sample with FPA, positive bias, (d) Annealed sample with FPA, negative bias, (e) Annealed sample with n-HPA, positive bias, (f) Annealed sample with n-HPA, negative bias.

FIG. 3. $\Delta V_{ON}$ shift vs positive and negative bias stress time (s) for IGZO TFTs with and without n-HPA and FPA. For annealed devices without SAMs and for a n-HPA passivated surface, a fit of the stretched exponential model (dashed lines) is shown. For devices with a FPA passivated surface, a linear fit (dashed lines) is shown.

FIG. 4. Surface sensitive O 1s XPS data obtained from blanket as deposited IGZO films and the same film during an *in-situ* anneal to 300 °C.

FIG. 5. O 1s XPS data obtained from blanket IGZO films that have been annealed and treated with n-HPA or FPA.

FIG. 6. C 1s XPS data obtained from blanket IGZO films that have been annealed and treated with n-HPA or FPA.

Figure 1:

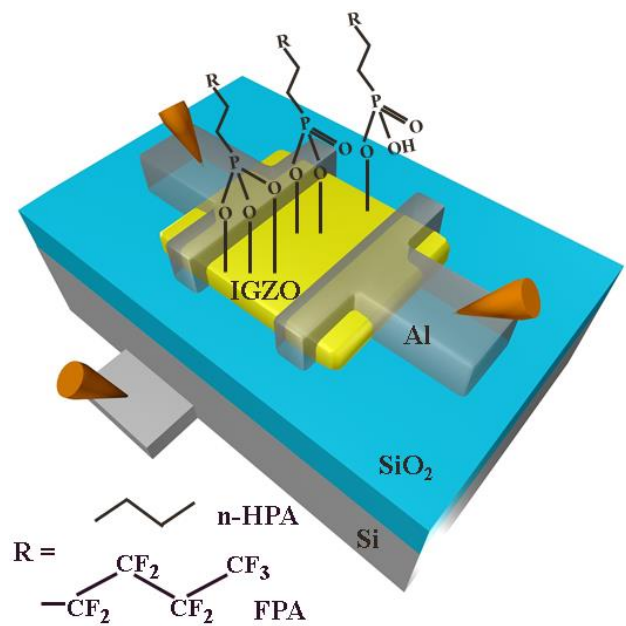

Figure 2:

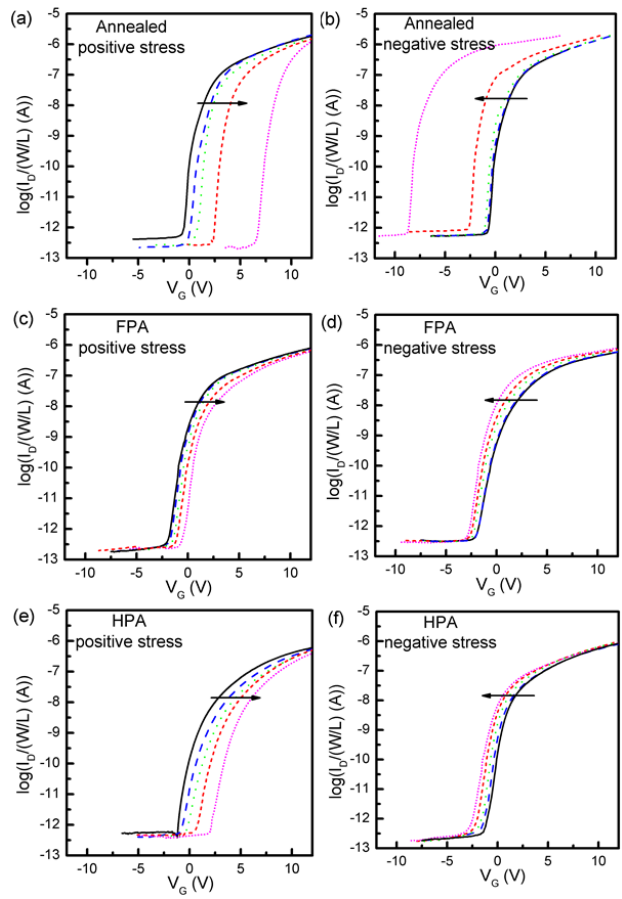

Figure 3:

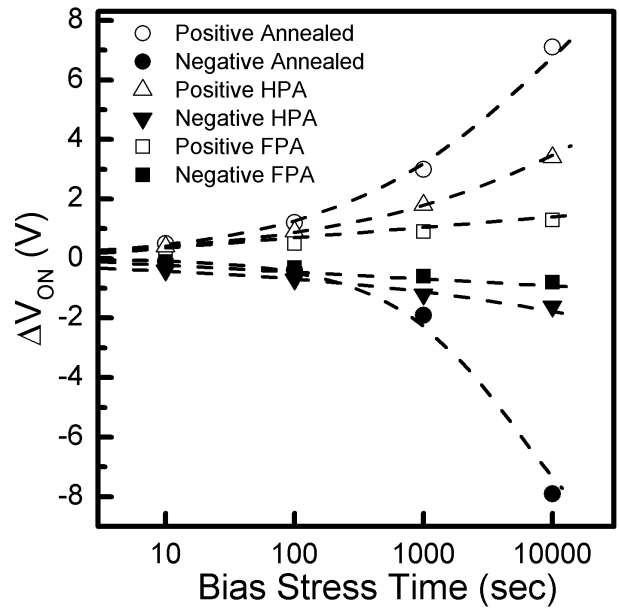

Figure 4:

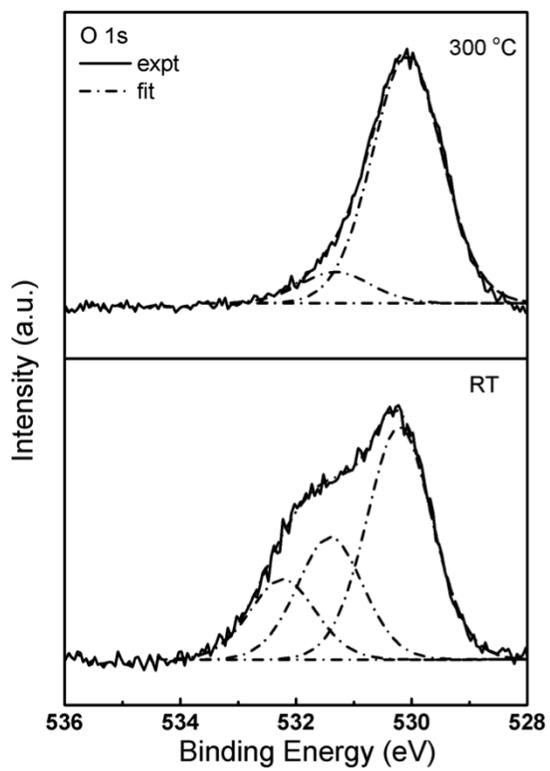

Figure 5:

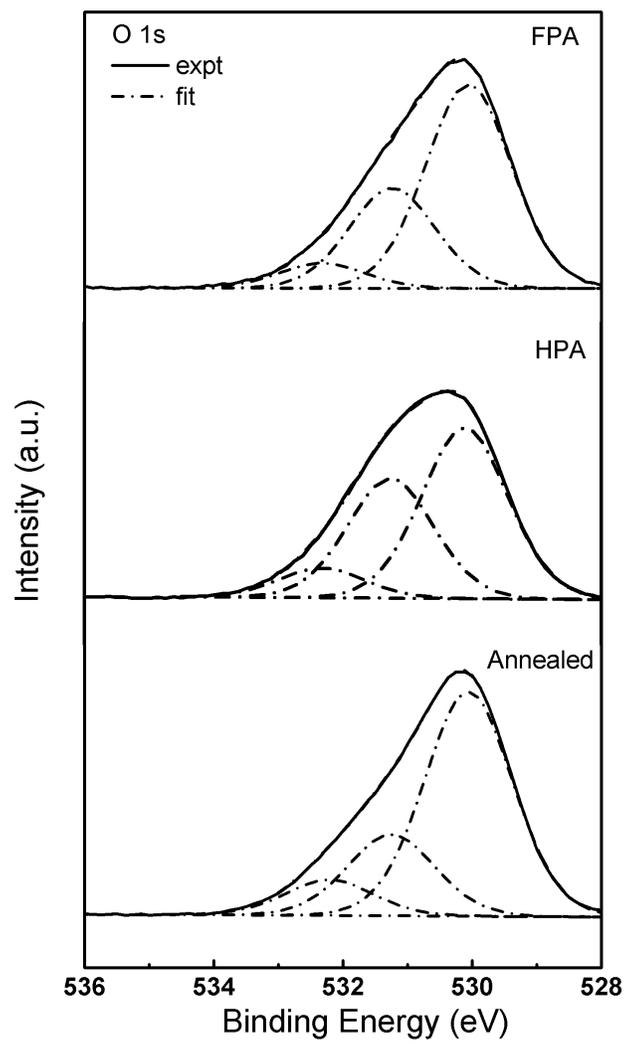

Figure 6:

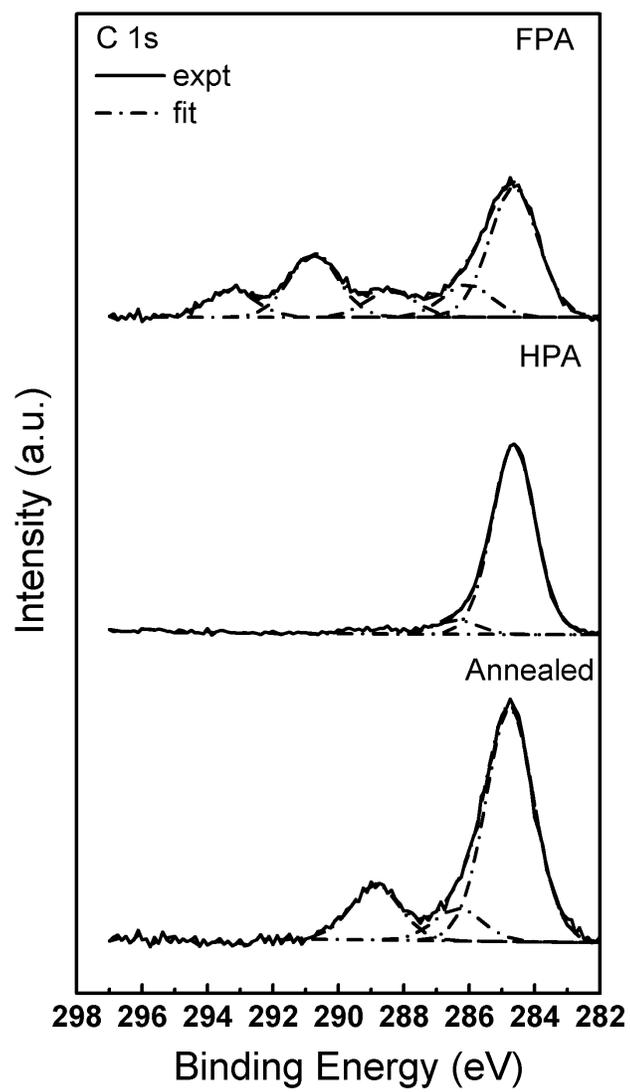